\documentclass[prd,preprint,nofootinbib,amsmath,amssymb]{revtex4-1}

\usepackage{amsmath, amssymb, mathtools}
\usepackage{bm}
\usepackage{mathrsfs}
\usepackage[colorlinks=true,linkcolor=blue,citecolor=blue]{hyperref}
\usepackage{cleveref}
\usepackage[utf8]{inputenc}

\newcommand{\be}{\begin{equation}}
\newcommand{\ee}{\end{equation}}

\newcommand{\p}{\partial}

\usepackage{color}

\newcommand{\ads}{$AdS_2$}

\usepackage{breqn}
\makeatletter
\let\cat@comma@active\@empty
\makeatother

\begin{document}

\title{\boldmath Scattering of gravitational and electromagnetic waves \\ 
			 	off $AdS_2\times S^2$ in extreme Reissner-Nordstrom}

\author{Achilleas P. Porfyriadis}
\affiliation{Department of Physics, UCSB, Santa Barbara, CA 93106, USA}


\begin{abstract}
	
The direct product of two-dimensional anti-de Sitter spacetime with a two-sphere is an exact solution of four-dimensional Einstein-Maxwell theory without a cosmological constant. In this paper, we analytically solve the coupled gravitational and electromagnetic perturbation equations of $AdS_2\times S^2$ in Einstein-Maxwell theory. On the other hand, $AdS_2\times S^2$ also describes the near-horizon region of the extreme Reissner-Nordstrom (ERN) black hole. We therefore also solve the connection problem: we show how the $AdS_2\times S^2$ perturbation equations arise from an appropriate near-horizon approximation of the corresponding equations for the ERN and then, using matched asymptotic expansions, we analytically extend the $AdS_2\times S^2$ solutions away from the near-horizon region connecting them with solutions in the far asymptotically flat region. From the point of view of ERN our results may be thought of as computing the classical scattering matrix for gravitational and electromagnetic waves which probe the near-horizon region of the black hole.

\end{abstract}

\maketitle


\section{Introduction}

$AdS_2$ is currently the focus of  much activity in the context of two-dimensional quantum gravity in asymptotically anti-de Sitter spacetimes; see e.g. \cite{Sarosi:2017ykf,Maldacena:2018lmt,Harlow:2018tqv} and references therein.
At the same time, $AdS_2$ has two major advantages over its higher dimensional counterparts in the AdS/CFT correspondence. The first advantage is that a direct product with a two-sphere produces an exact solution of four-dimensional pure Einstein-Maxwell theory without a cosmological constant, the so-called Bertotti-Robinson universe \cite{Bertotti:1959pf,Robinson:1959ev}. The second advantage is that the same solution also describes the near-horizon geometry of the asymptotically flat extreme Reissner-Nordstrom black hole; see e.g. \cite{Maldacena:1998uz}. This situates $AdS_2\times S^2$ in a privileged position for extending results produced in AdS/CFT investigations to the realm of four-dimensional classical and quantum gravity in asymptotically flat spacetimes. However, any such attempt requires that one first solves the \emph{connection problem}: one needs to know how to extend near-horizon anti-de Sitter solutions to the far asymptotically flat region. 

In this paper, we solve the connection problem for coupled gravitational and electromagnetic perturbations of $AdS_2\times S^2$ in Einstein-Maxwell theory. We show how the $AdS_2\times S^2$ perturbation equations arise from an appropriate near-horizon approximation of the corresponding equations for the full extreme Reissner-Nordstrom (ERN). We obtain analytically the exact solution to the $AdS_2\times S^2$ perturbation equations. We then find exact analytic solutions for two more approximations of the full ERN perturbation equations: an intermediate-region approximation whose solution is given by static perturbations and a far-region approximation whose solution gives the asymptotic behavior of the full ERN perturbations near infinity. Using the method of matched asymptotic expansions, we then glue the three solutions together and obtain the desired answer to our connection problem.

A reduction of the Reissner-Nordstrom perturbation equations to a set of two second-order radial differential equations may be found in Chapter 5 of the monograph \cite{Chandrasekhar:1985kt} and references therein; for a generalization to higher dimensions see \cite{Ishibashi:2011ws}. In general, these radial equations can only be solved numerically.
The method of matched asymptotic expansions has been used in the context of wave equations of (near-)extreme rotating black holes as early as \cite{Starobinsky:1973aij,Starobinskil:1974nkd,Teukolsky:1974yv}; for an application to the geodesic equations see \cite{Porfyriadis:2016gwb}.
Previous work indicating that ERN perturbations are at least partially amenable to analytic treatment includes \cite{Fabbri:1977ux,Crispino:2010fd}.

This paper is organized as follows. 
Section \ref{sec: review ERN, AdS2xS2, RWZ} contains a telegraphic review of Reissner-Nordstrom, its extreme near-horizon limit, the decoupled $AdS_2\times S^2$ geometry, and the Regge-Wheeler-Zerilli gauge for the pertubation equations. 
In Section \ref{sec: reduction of perturbation eqs} we present a reduction of the ERN perturbation equations to a single fourth-order radial differential equation. This reduction is different from the standard one which produces two second-order equations, but it contains the same information. 
In Section \ref{sec: main} we begin by setting up the matched asymptotic expansions for the full ERN radial equation. We then derive the corresponding $AdS_2\times S^2$ equation as a near-horizon approximation of the ERN equation and solve it analytically. Finally, we solve the connection problem by finding analytic solutions to intermediate and far region approximations of the full ERN equation and matching. 
Section \ref{sec: in and up solutions} specializes the connection formulas to the basis of solutions which are purely ingoing at the horizon in $AdS_2\times S^2$ and purely outgoing near infinity in ERN.

The case of near-extreme Reissner-Nordstrom and near-$AdS_2\times S^2$ is treated, to leading order in the deviation from extremality, in the companion paper \cite{Porfyriadis:2018jlw}.

\section{\boldmath Review of ERN, $AdS_2\times S^2$, and Regge-Wheeler-Zerilli gauge}\label{sec: review ERN, AdS2xS2, RWZ}

In this paper, we are concerned with the Einstein-Maxwell equations in four dimensions:\footnote{$G=c=1$}
\be\label{Einstein-Maxwell}
R_{\mu\nu}=8\pi T_{\mu\nu}\,,\qquad \nabla^\mu F_{\mu\nu}=0\,,
\ee
with $4\pi T_{\mu\nu}=F_{\mu\rho}F_\nu^{~\rho}-{1\over 4}g_{\mu\nu}F_{\rho\sigma}F^{\rho\sigma}$, and $F_{\mu\nu}=\p_\mu A_\nu-\p_\nu A_\mu$. Linearizing the equations around any solution we get
\begin{align}
&\nabla^\sigma\nabla_\mu h_{\sigma\nu}+\nabla^\sigma\nabla_\nu h_{\sigma\mu}-\Box h_{\mu\nu}-\nabla_\mu\nabla_\nu h \label{linearized Einstein} \\
&\quad = 2g_{\mu\nu}F_{\rho\lambda}F_\sigma^{~\lambda}h^{\rho\sigma}-4F_{\mu\rho}F_{\nu\sigma}h^{\rho\sigma}-F_{\rho\sigma}F^{\rho\sigma}h_{\mu\nu}+4F_\mu^{~\lambda}f_{\nu\lambda}+4F_\nu^{~\lambda}f_{\mu\lambda}-2g_{\mu\nu}F^{\rho\sigma}f_{\rho\sigma}\,, \notag \\
& \nabla^\mu f_{\mu\nu}-h^{\mu\sigma}\nabla_\sigma F_{\mu\nu}-\nabla_\mu h^{\mu\sigma} F_{\sigma\nu}-F^{\mu\sigma}\nabla_\mu h_{\sigma\nu}+\tfrac{1}{2}(\nabla^\mu h)F_{\mu\nu}=0\,, \label{linearized Maxwell}
\end{align}
where $h_{\mu\nu}\equiv\delta g_{\mu\nu}\,, a_\mu\equiv\delta A_\mu\,,f_{\mu\nu}\equiv\p_\mu a_\nu-\p_\nu a_\mu=\delta F_{\mu\nu}$ and $h\equiv g^{\mu\nu}h_{\mu\nu}$.

The Reissner-Nordstrom solution of the Einstein-Maxwell equations \eqref{Einstein-Maxwell} reads:
\be\label{RN hatted metric}
ds^2=-\left(1-{2M\over\hat{r}}+{Q^2\over\hat{r}^2}\right)d\hat{t}^2+\left(1-{2M\over\hat{r}}+{Q^2\over\hat{r}^2}\right)^{-1}d\hat{r}^2+\hat{r}^2d\Omega^2\,,\qquad \hat{A}_{\hat{t}}=-{Q\over \hat{r}}\,.
\ee
Extreme Reissner-Nordstrom (ERN) is given by $Q=M$:
\be\label{ERN hatted metric}
ds^2=-\left(1-{M\over\hat{r}}\right)^2d\hat{t}^2+\left(1-{M\over\hat{r}}\right)^{-2}d\hat{r}^2+\hat{r}^2d\Omega^2\,,\qquad \hat{A}_{\hat{t}}=-{M\over \hat{r}}\,.
\ee
The Bertotti-Robinson universe, which arises naturally as the near-horizon limit of ERN, may be obtained as follows. Make the coordinate and gauge transformation,
\be
r=\frac{\hat{r}-M}{M}\,,\quad t=\frac{\hat{t}}{M}\,,\qquad A=\hat{A}+d\hat{t}\,,
\ee
to obtain
\be\label{ERN lambda metric}
{1\over M^2} ds^2= -\left({r\over 1+r}\right)^2 dt^2 +\left({r\over 1+r}\right)^{-2} dr^2+(1+r)^2d\Omega^2\,,\qquad A_t=M{r\over 1+r}\,,
\ee
and then take the limit $r\ll 1$ to get:
\be\label{BR}
{1\over M^2} ds^2= -r^2 dt^2 + \frac{dr^2}{r^2}+d\Omega^2\,,\qquad A_t=Mr\,.
\ee
This solves the Einstein-Maxwell equations \eqref{Einstein-Maxwell} on its own. The metric in \eqref{BR} is $AdS_2\times S^2$.
From now on we set $M=1$.

Spherical symmetry allows one to separate the angular dependence in the linearized Einstein-Maxwell equations (\ref{linearized Einstein}--\ref{linearized Maxwell}) using scalar, vector, and tensor spherical harmonics labeled by $(l,m)$, belonging to two sectors of opposite parity: ``even'' parity $(-1)^l$ and ``odd'' parity $(-1)^{l+1}$. In this paper, we will consider the even parity sector only because we are primarily interested in perturbations to the $AdS_2$ part of the near-horizon metric and the odd sector does not allow any such perturbations.\footnote{Perturbations $h_{tt}, h_{tr}, h_{rr}$ transform as scalars on the sphere and therefore have the parity $(-1)^l$ of the standard (scalar) spherical harmonic functions $Y_{l,m}$.} Due to the spherical symmetry one may also set $m=0$ by an appropriate rotation. Time translation symmetry allows one to separate the time dependence as well using waves of definite energy $\omega$. Finally, we move to the Regge-Wheeler-Zerilli gauge thereby putting the following ansatz for the metric and gauge field perturbations \cite{Regge:1957td,Zerilli:1974ai}:
\begin{align}
h_{\mu\nu}=&
\begin{pmatrix}
	-Y(r)  & X(r)	& 0 	& 0 \\
	       & V(r)	& 0 	& 0 \\
	       & 	 	& K(r) 	& 0 \\
	       &  		&  		& K(r)\sin^2\theta
\end{pmatrix}
e^{i\omega t}\, Y_{l,0}(\theta,\phi) \,, \label{metric pertn ansatz}\\
a_\mu=&
\begin{pmatrix}
	\chi(r)  & \psi(r)	& 0 	& 0 
\end{pmatrix}	
e^{i\omega t}\, Y_{l,0}(\theta,\phi) \,. \label{gauge field pertn ansatz}
\end{align}
Note that modes with $l\geq 2$ may be treated in a unified manner while the cases $l=0$ and $l=1$ are special and need to treated separately.\footnote{Birkhoff's theorem implies that the only $l=0$ solutions are those corresponding to perturbations towards other Reissner-Nordstrom black holes parameterized by infinitesimal changes to the mass and charge parameters $\delta M,\delta Q$.} 
In this paper, we consider the $l\geq 2$ modes.


\section{Reduction of the perturbation equations}\label{sec: reduction of perturbation eqs}

Putting the ansatz (\ref{metric pertn ansatz}--\ref{gauge field pertn ansatz}) into the linearized Einstein-Maxwell equations (\ref{linearized Einstein}--\ref{linearized Maxwell}), on the background of the ERN \eqref{ERN lambda metric}, we may reduce them to a single fourth-order differential equation for $K$,
\be\label{Keqn}
a_4(r)K''''+a_3(r)K'''+a_2(r)K''+a_1(r)K'+a_0(r)K=0\,.
\ee
The expressions for the coefficients $a_i(r)$ may be found in App.~\ref{app: a's}. 
The solution to this differential equation fixes the remaining components of the gravitational and electromagnetic perturbation according to:
\begin{dgroup*}
\begin{dmath}\label{Y soln}
	Y=-\frac{1}{(1+r)^6 \left(l^2 (l+1)^2+4 (1+r)^2  \omega^2\right)}\times \left[l (l+1) (1+r)^2 r^4 K''-2 (1+r) r^2  \left(l (l+1) r^2-2 (1+r)^4 \omega^2\right)K' +\left(2 l (l+1) r^4-(1+r)^5 (l (l+1) (1+r)+4 r)\omega^2\right) K \right] \,,
\end{dmath}
\begin{dmath}\label{V soln}
V=-\frac{(1+r)^4}{r^4} Y \,,
\end{dmath}
\begin{dmath}\label{X soln}
	X=\frac{2 i \omega \left((1+r) r K'-(1+r) K-r^3 V\right)}{l(l+1)(1+r)r} \,,
\end{dmath}
\begin{dmath}\label{chi soln}
	\chi=\frac{r^3 (1+r) K'-2 r^3 K+r (1+r)^5 Y'-(1+r)^5 Y-r^4 (1+r) V+i\omega r (1+r)^5 X}{4 r (1+r)^3} \,,
\end{dmath}
\begin{dmath}\label{psi soln}
\psi=\frac{(1+r) \left(i \omega  (1+r) K+i \omega  r^2 (1+r) V-2 r X-r^2 (1+r) X'\right)}{4 r^2} \,.
\end{dmath}
\end{dgroup*}

The reduction described above is not the standard Regge-Wheeler-Zerilli one, who reduce to two second-order equations, but it contains the same information.
The reader may wonder what is the reduction path that leads to the above results. If we denote by $\mathcal{E}_{\mu\nu}$ and $\mathcal{M}_{\mu}$ the linearized Einstein \eqref{linearized Einstein} and Maxwell \eqref{linearized Maxwell} equations, respectively, then the reduction path may be described as follows. Begin by solving the equations $\mathcal{E}_{tr}, \mathcal{E}_{t\theta}, \mathcal{E}_{r\theta}, \mathcal{E}_{\phi\phi}-\sin^2\theta\, \mathcal{E}_{\theta\theta}$ which are algebraic equations for $X, \psi, \chi, V$, respectively. This gives rise to (\ref{V soln}--\ref{psi soln}). Then, using (\ref{V soln}--\ref{psi soln}), equation $\mathcal{E}_{tt}$ becomes algebraic for $Y$ and yields \eqref{Y soln}. At this stage, given (\ref{Y soln}--\ref{psi soln}), one is left with equations $\mathcal{E}_{\theta\theta}, \mathcal{M}_{t}, \mathcal{M}_{r}, \mathcal{M}_{\theta}$ for $K$ (the remaining equations vanish identically). However, one finds that $\mathcal{M}_{r}, \mathcal{M}_{\theta}$ are trivial multiples of $\mathcal{E}_{\theta\theta}$ while $\mathcal{M}_{t}$ is a linear combination of $\mathcal{E}_{\theta\theta}$ and its derivative. This leaves only equation $\mathcal{E}_{\theta\theta}$ for $K$, which is equation \eqref{Keqn}.

\section{\boldmath $AdS_2\times S^2$ answers and the connection problem solved}\label{sec: main}

Consider the scattering of gravitational and electromagnetic waves in the full ERN spacetime. 
Generic modes do not survive in the near-horizon limit. 
The modes which do survive and solve the $AdS_2\times S^2$ perturbation equations are low energy modes:
\be\label{low energy regime}
\omega\ll 1\,.
\ee
In this regime, to leading order in $\omega$, we may solve the scattering problem in the full ERN analytically using the method of matched asymptotic expansions as follows. Change variables in equation \eqref{Keqn} according to
\be\label{KtoH}
K(r)=\left(\frac{r}{1+r}\right)^2 H(r) \,,
\ee
and solve the resulting equation for $H$,
\be\label{Heqn}
b_4(r)H''''+b_3(r)H'''+b_2(r)H''+b_1(r)H'+b_0(r)H=0\,,
\ee
by dividing the spacetime into three regions:
\begin{align}
&\textrm{Near:}\qquad\qquad\quad \,r\ll 1 \label{near region}\\
&\textrm{Static:}\qquad\, \omega\ll \,r\ll 1/\omega\label{static region}\\
&\textrm{Far:}\qquad\quad\,\, 1\ll\,r \label{far region}
\end{align}
The intermediate Static region is overlapping with both the Near and the Far regions. Therefore, the Static solution may be used to connect the Near and Far solutions.

In the following, we will denote by $b_i^{n}(r), b_i^{s}(r), \text{and}\, b_i^{f}(r)$ the coefficients of Eq.~\eqref{Heqn}, in the Near, Static, and Far regions, respectively.

\subsection{Near region}\label{subsection: Near region}
In the Near region, which corresponds to the \ads$\times S^2$ throat, Eq.~\eqref{Heqn} reduces to
\begin{align*}
&b_4^{n}(r)=r^8 \\
&b_3^{n}(r)=16 r^7 \\
&b_2^{n}(r)=-2 \left(l^2+l-36\right) r^6+2 r^4 \omega^2 \\
&b_1^{n}(r)=-12 \left(l^2+l-8\right) r^5+8 r^3 \omega^2 \\
&b_0^{n}(r)=(l-3) (l-1) (l+2) (l+4) r^4 -2\left(l^2+l-2\right)r^2\omega^2+\omega^4 
\end{align*}
and the solution is given by
\be\label{Hnear soln}
H^{n}(r)=
C_1^n \, r^{-3}h_{-l-2}^{(1)}(\omega/r)+
C_2^n \,  r^{-3}h_{-l-2}^{(2)}(\omega/r)+
C_3^n \, r^{-3}h_{-l}^{(1)}(\omega/r)+
C_4^n \, r^{-3}h_{-l}^{(2)}(\omega/r)\,.
\ee
Here $h_{m}^{(1)},h_{m}^{(2)}$ are the spherical Hankel functions of the first and second kind, respectively.\footnote{Recall that for $m=0,1,2,\ldots$ we have: 
	\begin{align*}
	h_{m}^{(1)}(z)=i^{-m-1}{e^{iz}\over z}\sum_{k=0}^{m}(-1)^k\frac{(m+k)!}{k!(m-k)!}\frac{1}{(2iz)^k}\,,\\
	h_{m}^{(2)}(z)=i^{m+1}{e^{-iz}\over z}\sum_{k=0}^{m}(-1)^k\frac{(m+k)!}{k!(m-k)!}\frac{1}{(2iz)^k}\,,
	\end{align*}
	while $h_{-m-1}^{(1)}(z)=i(-1)^m h_{m}^{(1)}(z)$ and $h_{-m-1}^{(2)}(z)=-i(-1)^m h_{m}^{(2)}(z)$.
}

This Near solution is an exact solution with respect to the Bertotti-Robinson universe. That is to say, using the perturbation ansatz (\ref{metric pertn ansatz}--\ref{gauge field pertn ansatz}) on the background of the Bertotti-Robinson universe \eqref{BR}, Eq.~\eqref{Hnear soln} gives the exact solution of the linearized Einstein-Maxwell equations (\ref{linearized Einstein}--\ref{linearized Maxwell}), provided that Eqs.~(\ref{Y soln}--\ref{psi soln}) are replaced by
\begin{align*}
&\qquad\, Y=-\frac{r^4 K''-\omega ^2 K}{l (l+1)}\,,\quad 	V=-\frac{1}{r^4} Y\,,\quad 		X=-\frac{2 i \omega  \left(K-r K'\right)}{l (l+1) r}\,,\\
&\chi=-\frac{-r^3 K'+Y-r Y'+r^4 V-i \omega  r X}{4 r}\,,\quad \psi=\frac{i \omega  K+i \omega  r^2 V-2 r X-r^2 X'}{4 r^2}\,,
\end{align*}
and Eq.~\eqref{KtoH} is replaced by $K(r)=r^2 H(r)$.

\subsection{Static region} 
In the Static region, which corresponds to setting $\omega=0$, Eq.~\eqref{Heqn} reduces to
\begin{align*}
&b_4^{s}(r)=r^4 (1+r)^4\\
&b_3^{s}(r)=2 r^3 (1+r)^3 (8+r)\\
&b_2^{s}(r)=-2 r^2 (1+r)^2 \left(l (l+1) (1+r)^2-36+10 r\right)\\
&b_1^{s}(r)=-2 r (1+r) \left(l (l+1) (1+r)^2 (6-r)-48+94 r-20 r^2\right)\\
&b_0^{s}(r)=l (l+1) (1+r)^2 \left(l (l+1) (1+r)^2-2 (1-r) (7-r)\right)+8 \left(3-26 r+29 r^2-5 r^3\right)
\end{align*}
and the solution is given by
\begin{align}\label{Hstatic soln}
H^{s}(r)=(1+r)^3 
\left[C_1^s \, r^{l-1} +
C_2^s \, r^{l-3} (l-r+2 l r) +
C_3^s \, r^{-l-2} +
C_4^s \, r^{-l-4} (l+1+3 r+2 l r)\right]\,.
\end{align}

\subsection{Far region} 
In the Far region, which includes asymptotically flat infinity, Eq.~\eqref{Heqn} reduces to
\begin{align*}
&b_4^{f}(r)=r^4 \left(l^2 (l+1)^2+4 r^2 \omega ^2\right)^2\\
&b_3^{f}(r)=2r^3 \left(l^4 (l+1)^4-16 r^4 \omega ^4\right)\\
&b_2^{f}(r)=-2 r^2 \left[l^5 (l+1)^5-l^2 (l+1)^2 
\left(l^2 (l+1)^2-4 l (l+1)-4\right) r^2 \omega^2 \right. \\ 
&\qquad\qquad\qquad\, \left. -8 \left(l^2 (l+1)^2+6\right) r^4 \omega^4
-16 r^6 \omega^6\right]\\
&b_1^{f}(r)=2 r \left[l^5 (l+1)^5+l^2 (l+1)^2 \left(l^2 (l+1)^2+20 l (l+1)+8\right) r^2 \omega ^2-96 r^4 \omega ^4-16 r^6 \omega ^6\right]\\
&b_0^{f}(r)=l^6 (l+1)^6-2 l^5 (l+1)^5
-2 l^2 (l+1)^2 \left(l^3 (l+1)^3-2 l^2 (l+1)^2+20 l (l+1)+8\right) r^2 \omega^2  \\
&\qquad\quad +\left(l^4 (l+1)^4-8 l^3 (l+1)^3 
+56 l^2 (l+1)^2+192\right)r^4\omega^4  \\
&\qquad\quad+8 \left(l^2 (l+1)^2+12\right) r^6 \omega ^6+16 r^8 \omega ^8 
\end{align*}
and the solution is given by
\begin{align}\label{Hfar soln}
H^{f}(r)=
C_1^f &\,  r \omega  
\left(
(l+1) (l+2) r \omega  h_{l+2}^{(1)}(r \omega )
-\left((l+1) (l+2) (2 l+3) -2 r^2 \omega^2\right) h_{l+1}^{(1)}(r \omega)
\right) \notag\\+
C_2^f &\, r \omega  
\left(
(l+1) (l+2) r \omega  h_{l+2}^{(2)}(r \omega )
-\left( (l+1) (l+2) (2 l+3) -2 r^2 \omega^2\right) h_{l+1}^{(2)}(r \omega)
\right)\notag\\+
C_3^f &\, \left(
r \omega  \left(l (l+1)^3 (l+2)-2 \left(l^2+l+2\right) r^2 \omega ^2\right) h_{l+2}^{(1)}(r \omega )\right.\notag\\
&\quad\left.-(l+2) \left(l (l+1)^3 (2 l+3)-(l+3)
\left(l^2+l+2\right) r^2 \omega ^2\right) h_{l+1}^{(1)}(r \omega )
\right)\notag\\+
C_4^f &\, \left(r \omega  \left(l (l+1)^3 (l+2)-2 \left(l^2+l+2\right) r^2 \omega
^2\right) h_{l+2}^{(2)}(r \omega )\right.\notag\\
&\quad\left.-(l+2) \left(l (l+1)^3 (2 l+3)-(l+3) \left(l^2+l+2\right) r^2 \omega ^2\right) h_{l+1}^{(2)}(r \omega )\right)\,.
\end{align}

\subsection{Overlap regions, matching, and the solution to the connection problem}

The Static region overlaps with the Near region in 
\be \label{near-static overlap region}
\textrm{Near-Static overlap:}\qquad\quad \,\omega\ll \,r\ll 1 
\ee
where Eq.~\eqref{Heqn} reduces to
\be
r^4 H''''+16 r^3 H'''-2 \left(l^2+l-36\right) r^2 H''-12 \left(l^2+l-8\right) r H'+(l-3) (l-1) (l+2) (l+4)H=0\notag
\ee
and the solution is
\be\label{HstaticMATCHnear soln}
H^{ns}(r)=C_1^{ns} r^{l-1}+C_2^{ns} r^{l-3}+C_3^{ns} r^{-l-2}+C_4^{ns} r^{-l-4}\,.
\ee
Expanding the Near solution \eqref{Hnear soln} around infinity and the Static solution \eqref{Hstatic soln} around the origin, one may obtain linear relations between the $C_i^n$'s and $C_i^s$'s by matching the coefficients of the two expansions with the Near-Static solution \eqref{HstaticMATCHnear soln}.

The Static region overlaps with the Far region in 
\be \label{far-static overlap region}
\textrm{Far-Static overlap:}\qquad\quad \,1\ll \,r\ll 1/\omega 
\ee
where Eq.~\eqref{Heqn} reduces to
\be
r^4 H''''+ 2 r^3 H''' -2 l (l+1) r^2 H'' +2 l (l+1) r H'+ (l-1) l (l+1) (l+2)H=0\notag
\ee
and the solution is
\be\label{HstaticMATCHfar soln}
H^{fs}(r)=C_1^{fs} r^{l+2}+C_2^{fs} r^{l+1}+C_3^{fs} r^{-l+1}+C_4^{fs} r^{-l}\,.
\ee
Expanding the Far solution \eqref{Hfar soln} around the origin and the Static solution \eqref{Hstatic soln} around infinity, one may obtain linear relations between the $C_i^f$'s and $C_i^s$'s by matching the coefficients of the two expansions with the Far-Static solution \eqref{HstaticMATCHfar soln}.

Eliminating the $C_i^s$'s from the matching equations outlined above, we  obtain the following linear relation between the $C_i^f$'s and $C_i^n$'s:
\begin{align}
&\begin{aligned}
C_{12}^{n+}
=\frac{(-1)^{l+1}}{2^{2 l+2}\pi}\frac{(l+1) (l+2) }{ 2 l-1}
\Gamma \left(-l-\tfrac{1}{2}\right)^2  \omega^{2 l+3}
\left( (25 l-8) \omega \,C_{12}^{f+} -3  l(l+1)^2 (3 l-1) \,C_{34}^{f+} \right) \notag
\end{aligned} 
\\
&\begin{aligned} \label{scattering eqns soln}
&C_{12}^{n-}
=\frac{(-1)^l 2^{2l}}{\pi} (l-1) l (l+1) (2l+1) 
\Gamma \left(l+\tfrac{1}{2}\right)^2  \omega^{-2 l-1}
\left(3\omega \,C_{12}^{f-} + l^2 (l+1) \,C_{34}^{f-} \right)
\\
&C_{34}^{n+}
=\frac{(-1)^{l+1}}{2^{2l+2}\pi}  l(l+1) (l+2) (2l+1)  
\Gamma \left(-l-\tfrac{1}{2}\right)^2  \omega^{2 l+1}
\left(3 \omega \,C_{12}^{f+} - l (l+1)^2 \,C_{34}^{f+} \right) 
\end{aligned}
\\
&\begin{aligned}
C_{34}^{n-}
=\frac{(-1)^l 2^{2l}}{\pi}\frac{ (l-1) l }{2 l+3} 
\Gamma \left(l+\tfrac{1}{2}\right)^2  \omega^{-2 l+1}
\left((25l+33)\omega \,C_{12}^{f-} +3 l^2(l+1)(3l+4) \,C_{34}^{f-} \right) \notag
\end{aligned}
\end{align}
where we have defined
\begin{align}
\begin{aligned}
C_{12}^{n\pm}=C_{1}^{n}\pm C_{2}^{n}\,,\quad C_{34}^{n\pm}=C_{3}^{n}\pm C_{4}^{n}\,,\\
C_{12}^{f\pm}=C_{1}^{f}\pm C_{2}^{f}\,,\quad C_{34}^{f\pm}=C_{3}^{f}\pm C_{4}^{f}\,.
\end{aligned}
\end{align}
This is the main result of the paper and completes the solution to our connection problem. The connection formulas \eqref{scattering eqns soln} may also be thought of as computing the classical scattering matrix for gravitational and electromagnetic waves which probe the near-horizon region of extreme Reissner-Nordstrom.

\section{A useful basis of solutions}\label{sec: in and up solutions}

A  useful basis of solutions for the full ERN equation \eqref{Heqn} consists of two solutions $H_1, H_2$ which are purely ingoing near the horizon and two solutions $H_3, H_4$ which are purely outgoing near infinity. Given that for $z\to\infty$ we have $h_{m}^{(1)}(z)\sim i^{-m-1} z^{-1} e^{i z}$ and $h_{m}^{(2)}(z)\sim i^{m+1} z^{-1} e^{-i z}$ this basis is defined by:
\begin{align}
H_1 \quad \textrm{is the solution with:}\quad C^n_2=1\,,C^n_1=C^n_3=C^n_4=0\,,\\
H_2\quad \textrm{is the solution with:}\quad C^n_4=1\,,C^n_1=C^n_3=C^n_2=0\,,\\
H_3 \quad \textrm{is the solution with:}\quad C^f_2=1\,,C^f_1=C^f_3=C^f_4=0\,,\\
H_4\quad \textrm{is the solution with:}\quad C^f_4=1\,,C^f_1=C^f_3=C^f_2=0\,.
\end{align}
From the connection formulas \eqref{scattering eqns soln} we find that:

$H_1$ has the Far amplitudes
\begin{align}\label{H1 far amplitudes}
\begin{aligned}
C^f_1=C^f_2=&
{(-1)^l\over \pi}\frac{2^{2 l+1} \Gamma \left(l+\frac{3}{2}\right)^2}{(l+1) (l+2) \omega^{2l+4}}\,,
\\
C^f_3=C^f_4=&
{(-1)^l\over \pi}\frac{3\cdot 2^{2 l+1} \Gamma \left(l+\frac{3}{2}\right)^2}{l(l+1)^3 (l+2) \omega^{2l+3}}\,,
\end{aligned}
\end{align}

$H_2$ has the Far amplitudes
\begin{align}\label{H2 far amplitudes}
\begin{aligned}
C^f_1=C^f_2=&
{(-1)^{l+1}\over \pi}\frac{3\cdot 2^{2 l+1} (3 l-1) 
\Gamma \left(l+\frac{3}{2}\right)^2}{l (l+1) (l+2) (2 l-1) (2 l+1) \omega^{2l+2}}\,,
\\
C^f_3=C^f_4=&
{(-1)^{l+1}\over \pi}\frac{2^{2 l+1} (25 l-8)  
\Gamma \left(l+\frac{3}{2}\right)^2}{l^2 (l+1)^3 (l+2) (2 l-1) (2 l+1) \omega^{2l+1}}\,,
\end{aligned}
\end{align}

$H_3$ has the Near amplitudes
\begin{align}
\begin{aligned}
C^n_1=-C^n_2=&
{(-1)^{l+1}\over \pi}\frac{3\cdot 2^{2 l-1} (l-1) l (l+1) (2 l+1) 
\Gamma \left(l+\frac{1}{2}\right)^2}{\omega^{2l}}\,,
\\
C^n_3=-C^n_4=&
{(-1)^{l+1}\over \pi}\frac{2^{2 l-1} (l-1) l (25 l+33)  
\Gamma \left(l+\frac{1}{2}\right)^2}{(2 l+3)\omega ^{2 l-2}}\,,
\end{aligned}
\end{align}

$H_4$ has the Near amplitudes
\begin{align}
\begin{aligned}
C^n_1=-C^n_2=&
{(-1)^{l+1}\over \pi}\frac{2^{2 l-1} (l-1) l^3 (l+1)^2 (2 l+1) 
\Gamma \left(l+\frac{1}{2}\right)^2}{\omega^{2l+1}}\,,
\\
C^n_3=-C^n_4=&
{(-1)^{l+1}\over \pi}\frac{3\cdot 2^{2 l-1} (l-1) l^3 (l+1) (3 l+4) 
\Gamma \left(l+\frac{1}{2}\right)^2}{(2 l+3)\omega ^{2 l-1}}\,.
\end{aligned}
\end{align}

A word of caution is in order here. 
$H_1, H_2$ and $H_3, H_4$ differ from the standard ``in'' and ``up'' solutions\footnote{The labels ``in'' and ``up'' for solutions purely ingoing at the horizon and purely outgoing at infinity, respectively, were introduced in \cite{Chrzanowski:1974nr} and have become standard in black hole perturbation theory \cite{Frolov:1998wf, Sasaki:2003xr}.} one would define for ERN by infinitely oscillating phases at the horizon and infinity, respectively. 
For example, an ``in'' solution in ERN is defined as one that behaves like $e^{i\omega r_*}$ near the horizon, where $r_*=-1/r+2\ln r+r$ is the ERN tortoise coordinate. On the other hand, the $AdS_2\times S^2$ tortoise coordinate is $r_*=-1/r$ so that the behavior of $H_1, H_2$ near the $AdS_2\times S^2$ horizon agrees, for sufficiently small $\omega$, with that of an ``in'' ERN solution everywhere near the horizon except precisely at the horizon where the two differ by the infinitely oscillating phase $e^{2i\omega\ln r}$.  
Similar comments apply to the behavior at infinity.

\section{Conclusion}\label{sec: conclusion}

In this paper we have analytically solved the coupled gravitational and electromagnetic perturbation equations of the Bertotti-Robinson universe. Moreover, using matched asymptotic expansions, we have shown how to analytically extend these solutions to the asymptotically flat region of extreme Reissner-Nordstrom. We expect this extension to prove useful for transferring results from studies of $AdS_2$ holography to the realm of four-dimensional classical and quantum gravity in asymptotically flat spacetimes.

In has been argued that $AdS_2$ has no finite energy excitations \cite{Maldacena:1998uz}. The reason is that, with anti-de Sitter boundary conditions, backreaction in $AdS_2$ is so strong that it destroys its asymptotics. The reader might therefore wonder how do the results of this paper fit in that picture. We note that it is possible to study the backreaction problem using the linearized perturbations of $AdS_2\times S^2$ that we have explicitly found in Section \ref{subsection: Near region}. The technique uses the so-called ``linearization stability constraints'' and we expect that, with anti-de Sitter boundary conditions, one would confirm that the solutions of Section \ref{subsection: Near region} should be discarded. Indeed, we expect that the only solution which obeys asymptotically $AdS_2\times S^2$ boundary conditions is diffeomorphic to $AdS_2\times S^2$ itself; see e.g. \cite{Galloway:2018dak}. However, the main focus of this paper is the connection with the asymptotically flat extreme Reissner-Nordstrom and in this context ``leaky boundary conditions'' at the location of the would-be $AdS_2\times S^2$ boundary are both allowed and desired. That is to say, energy is allowed to flow in and out of the anti-de Sitter boundary as needed in accordance with the boundary conditions imposed at asymptotically flat infinity.\footnote{For readers who are familiar with the NHEK geometry \cite{Bardeen:1999px} in the extreme Kerr throat, we note that the situation here is exactly analogous to the one in Kerr/CFT \cite{Guica:2008mu}. Linearization stability constraints imply that, up to diffeomorphisms and boundary gravitons, asymptotically NHEK boundary conditions are consistent only with trivial solutions $h_{\mu\nu}=0$ \cite{Amsel:2009ev, Dias:2009ex}. On the other hand, with astrophysically meaningful boundary conditions imposed at future null infinity of extreme Kerr, Refs.~\cite{Porfyriadis:2014fja,Hadar:2015xpa} analytically computed the gravitational wave signals from extreme-mass-ratio-inspirals in NHEK.} Consider for example the scattering problem set up by sending gravitational and electromagnetic waves from the asymptotically flat region in extreme Reissner-Nordstrom. The boundary conditions for such a problem are those of the solutions $H_1$ and $H_2$ in Section \ref{sec: in and up solutions}. In this scattering problem, gravitational and electromagnetic energy flows in from outside the $AdS_2\times S^2$ boundary, scatters off the horizon, and then flows out of the $AdS_2\times S^2$ boundary, with a ratio that is dictated by the amplitudes in Eqs.~\eqref{H1 far amplitudes} and \eqref{H2 far amplitudes}.

A model of two-dimensional anti-de Sitter gravity that takes into account backreaction is the Jackiw-Teitelboim theory \cite{Jackiw:1984je,Teitelboim:1983ux,Almheiri:2014cka} and it has been recently shown that the effective action which describes all gravitational dynamics in this theory is given by the Schwarzian derivative of an $AdS_2$ boundary reparametrization $t\to f(t)$ \cite{Maldacena:2016upp}. The Jackiw-Teitelboim theory describes the s-wave sector of higher dimensional theories with an $AdS_2$ throat geometry. In the context of this paper, this means that the Schwarzian action for a linearized reparametrization $t\to t+\epsilon(t)$ should capture the gravitational dynamics of the $l=0$ solutions of the linearized Einstein-Maxwell equations in extreme Reissner-Nordstrom. However, Birkhoff's theorem in four-dimensions implies that there are no nontrivial $l=0$ solutions other than linear perturbations towards other Reissner-Nordstrom black holes parametrized by small changes $\delta M,\delta Q$ of the mass and charge of the black hole. Indeed, it is not difficult to verify (see e.g. \cite{Nayak:2018qej}) that, up to the $SL(2,R)$ isometries of $AdS_2$, the only solution to the Schwarzian action expanded to second order in $\epsilon$ is given by $\epsilon(t)=\alpha t^3$ with $\alpha\propto \delta M-\delta Q$.

\acknowledgements

I am grateful to Sam Gralla and Gary Horowitz for useful conversations.  
This work is supported by NSF grant PHY-1504541

\appendix
\section{The master radial equation for ERN perturbations}\label{app: a's}

In Section \ref{sec: reduction of perturbation eqs} we presented a reduction of the ERN perturbation equations to the fourth-order differential equation \eqref{Keqn}.
The expressions for the coefficients $a_i(r)$ appearing in Eq.~\eqref{Keqn} are given by:

\begin{dgroup*}
	\begin{dmath*}
		a_4(r)=r^8 (1+r)^4 \left(l^2 (1+l)^2+4 (1+r)^2 \omega ^2\right)^2
	\end{dmath*}
	\begin{dmath*}
		a_3(r)=2 r^7 (1+r)^3 \left(l^4 (1+l)^4 (4+r)+32 l^2 (1+l)^2 (1+r)^2 \omega ^2 +16 (4-r) (1+r)^4 \omega ^4\right)
	\end{dmath*}
	\begin{dmath*}
		a_2(r)=2 r^4 (1+r)^2 \left[-l^4 (1+l)^4 r^2 \left(l (1+l) (1+r)^2+4 r-6\right)+l^2 (1+l)^2 (1+r)^2 \left(l^2 (1+l)^2 (1+r)^4-4 l (1+l) r^2 (1+r)^2+4 r^2 \left(12-16 r-r^2\right)\right) \omega^2+8 (1+r)^4 \left(l^2 (1+l)^2 (1+r)^4+6 r^2 \left(2-4 r+r^2\right)\right) \omega^4 +16 (1+r)^{10} \omega^6\right]
	\end{dmath*}
	\begin{dmath*}
		a_1(r)=2 r^4 (1+r) \left[-l^4 (1+l)^4 r \left(l (1+l) \left(2+3 r-r^3\right)+4 r (3-2 r)\right)+l^2 (1+l)^2 (1+r)^2 \left(l^2 (1+l)^2
		(1+r)^4-4 l (1+l) r (1+r)^2 (2-5 r)-8 r^2 \left(12-16 r-r^2\right)\right) \omega ^2-96 r^2 (1+r)^4 \left(2-4 r+r^2\right) \omega^4-16 (1+r)^{10} \omega ^6\right]
	\end{dmath*}
	\begin{dmath*}
		a_0(r)=l^4 (1+l)^4 r^4 \left(l^2 (1+l)^2 (1+r)^4-2 l (1+l) (1-r)^2 (1+r)^2+8 r^2 (3-2 r)\right)
		-2 l^2 (1+l)^2 r^2 (1+r)^2 \left(l^3(1+l)^3 (1+r)^6+2 l^2 (1+l)^2 (1-r) r (1+r)^4+4 l (1+l) r^2 (1+r)^2 \left(4-r+5 r^2\right)-8 r^4 \left(12-16 r-r^2\right)\right)	\omega ^2
		+(1+r)^4 \left(l^4 (1+l)^4 (1+r)^8-8 l^3 (1+l)^3 r^2 (1+r)^6+56 l^2 (1+l)^2 r^4 (1+r)^4-96 l (1+l) (1-r) r^4 (1+r)^2+192 r^6 \left(2-4 r+r^2\right)\right) \omega^4
		+8 (1+r)^{10} \left(l^2 (1+l)^2 (1+r)^4+4 r^3 (2+3 r)\right) \omega^6
		+16	(1+r)^{16} \omega ^8
	\end{dmath*}		
\end{dgroup*}


\begin{thebibliography}{99}
	

\bibitem{Sarosi:2017ykf} 
G.~Sárosi,
``AdS$_{2}$ holography and the SYK model,''
PoS Modave {\bf 2017}, 001 (2018)
doi:10.22323/1.323.0001
[arXiv:1711.08482 [hep-th]].

\bibitem{Maldacena:2018lmt} 
J.~Maldacena and X.~L.~Qi,
``Eternal traversable wormhole,''
arXiv:1804.00491 [hep-th].

\bibitem{Harlow:2018tqv} 
D.~Harlow and D.~Jafferis,
``The Factorization Problem in Jackiw-Teitelboim Gravity,''
arXiv:1804.01081 [hep-th].

\bibitem{Bertotti:1959pf} 
B.~Bertotti,
``Uniform electromagnetic field in the theory of general relativity,''
Phys.\ Rev.\  {\bf 116}, 1331 (1959).
doi:10.1103/PhysRev.116.1331

\bibitem{Robinson:1959ev} 
I.~Robinson,
``A Solution of the Maxwell-Einstein Equations,''
Bull.\ Acad.\ Pol.\ Sci.\ Ser.\ Sci.\ Math.\ Astron.\ Phys.\  {\bf 7}, 351 (1959).

\bibitem{Maldacena:1998uz} 
J.~M.~Maldacena, J.~Michelson and A.~Strominger,
``Anti-de Sitter fragmentation,''
JHEP {\bf 9902}, 011 (1999)
doi:10.1088/1126-6708/1999/02/011
[hep-th/9812073].

\bibitem{Chandrasekhar:1985kt} 
S.~Chandrasekhar,
``The mathematical theory of black holes,''
(Clarendon Press, Oxford, England, 1985).

\bibitem{Ishibashi:2011ws} 
A.~Ishibashi and H.~Kodama,
``Perturbations and Stability of Static Black Holes in Higher Dimensions,''
Prog.\ Theor.\ Phys.\ Suppl.\  {\bf 189}, 165 (2011)
doi:10.1143/PTPS.189.165
[arXiv:1103.6148 [hep-th]].

\bibitem{Starobinsky:1973aij} 
A.~A.~Starobinsky,
``Amplification of waves reflected from a rotating "black hole".,''
Sov.\ Phys.\ JETP {\bf 37}, no. 1, 28 (1973)
[Zh.\ Eksp.\ Teor.\ Fiz.\  {\bf 64}, 48 (1973)].

\bibitem{Starobinskil:1974nkd} 
A.~A.~Starobinsky and S.~M.~Churilov,
``Amplification of electromagnetic and gravitational waves scattered by a rotating "black hole",''
Sov.\ Phys.\ JETP {\bf 65}, no. 1, 1 (1974).

\bibitem{Teukolsky:1974yv} 
S.~A.~Teukolsky and W.~H.~Press,
``Perturbations of a rotating black hole. III - Interaction of the hole with gravitational and electromagnet ic radiation,''
Astrophys.\ J.\  {\bf 193}, 443 (1974).
doi:10.1086/153180

\bibitem{Porfyriadis:2016gwb} 
A.~P.~Porfyriadis, Y.~Shi and A.~Strominger,
``Photon Emission Near Extreme Kerr Black Holes,''
Phys.\ Rev.\ D {\bf 95}, no. 6, 064009 (2017)
doi:10.1103/PhysRevD.95.064009
[arXiv:1607.06028 [gr-qc]].


\bibitem{Fabbri:1977ux} 
R.~Fabbri,
``Electromagnetic and Gravitational Waves in the Background of a Reissner-Nordstrom Black Hole,''
Nuovo Cim.\ B {\bf 40}, 311 (1977).
doi:10.1007/BF02728215


\bibitem{Crispino:2010fd} 
L.~C.~B.~Crispino, A.~Higuchi and G.~E.~A.~Matsas,
``Low-frequency absorption cross section of the electromagnetic waves for the extreme Reissner-Nordstrom black holes in higher dimensions,''
Phys.\ Rev.\ D {\bf 82}, 124038 (2010)
doi:10.1103/PhysRevD.82.124038
[arXiv:1004.4018 [gr-qc]].




\bibitem{Porfyriadis:2018jlw} 
A.~P.~Porfyriadis,
``near-$AdS_2$ perturbations and the connection with near-extreme Reissner-Nordstrom,''
arXiv:1806.07097 [hep-th].


\bibitem{Regge:1957td} 
T.~Regge and J.~A.~Wheeler,
``Stability of a Schwarzschild singularity,''
Phys.\ Rev.\  {\bf 108}, 1063 (1957).
doi:10.1103/PhysRev.108.1063

\bibitem{Zerilli:1974ai} 
F.~J.~Zerilli,
``Perturbation analysis for gravitational and electromagnetic radiation in a reissner-nordstroem geometry,''
Phys.\ Rev.\ D {\bf 9}, 860 (1974).
doi:10.1103/PhysRevD.9.860

\bibitem{Chrzanowski:1974nr} 
P.~L.~Chrzanowski and C.~W.~Misner,
``Geodesic synchrotron radiation in the Kerr geometry by the method of asymptotically factorized Green's functions,''
Phys.\ Rev.\ D {\bf 10}, 1701 (1974).
doi:10.1103/PhysRevD.10.1701

\bibitem{Frolov:1998wf} 
V.~P.~Frolov and I.~D.~Novikov,
``Black hole physics: Basic concepts and new developments,''
(Kluwer Academic, Dodrecht, Netherlands, 1998)

\bibitem{Sasaki:2003xr} 
M.~Sasaki and H.~Tagoshi,
``Analytic black hole perturbation approach to gravitational radiation,''
Living Rev.\ Rel.\  {\bf 6}, 6 (2003)
doi:10.12942/lrr-2003-6
[gr-qc/0306120].

\bibitem{Galloway:2018dak} 
G.~J.~Galloway and M.~Graf,
``Rigidity of asymptotically $AdS_2 \times S^2$ spacetimes,''
arXiv:1803.10529 [gr-qc].

\bibitem{Bardeen:1999px} 
J.~M.~Bardeen and G.~T.~Horowitz,
``The Extreme Kerr throat geometry: A Vacuum analog of AdS(2) x S(2),''
Phys.\ Rev.\ D {\bf 60}, 104030 (1999)
doi:10.1103/PhysRevD.60.104030
[hep-th/9905099].

\bibitem{Guica:2008mu} 
M.~Guica, T.~Hartman, W.~Song and A.~Strominger,
``The Kerr/CFT Correspondence,''
Phys.\ Rev.\ D {\bf 80}, 124008 (2009)
doi:10.1103/PhysRevD.80.124008
[arXiv:0809.4266 [hep-th]].

\bibitem{Amsel:2009ev} 
A.~J.~Amsel, G.~T.~Horowitz, D.~Marolf and M.~M.~Roberts,
``No Dynamics in the Extremal Kerr Throat,''
JHEP {\bf 0909}, 044 (2009)
doi:10.1088/1126-6708/2009/09/044
[arXiv:0906.2376 [hep-th]].

\bibitem{Dias:2009ex} 
O.~J.~C.~Dias, H.~S.~Reall and J.~E.~Santos,
``Kerr-CFT and gravitational perturbations,''
JHEP {\bf 0908}, 101 (2009)
doi:10.1088/1126-6708/2009/08/101
[arXiv:0906.2380 [hep-th]].

\bibitem{Porfyriadis:2014fja} 
A.~P.~Porfyriadis and A.~Strominger,
``Gravity waves from the Kerr/CFT correspondence,''
Phys.\ Rev.\ D {\bf 90}, no. 4, 044038 (2014)
doi:10.1103/PhysRevD.90.044038
[arXiv:1401.3746 [hep-th]].


\bibitem{Hadar:2015xpa} 
S.~Hadar, A.~P.~Porfyriadis and A.~Strominger,
``Fast plunges into Kerr black holes,''
JHEP {\bf 1507}, 078 (2015)
doi:10.1007/JHEP07(2015)078
[arXiv:1504.07650 [hep-th]].



\bibitem{Jackiw:1984je} 
R.~Jackiw,
``Lower Dimensional Gravity,''
Nucl.\ Phys.\ B {\bf 252}, 343 (1985).
doi:10.1016/0550-3213(85)90448-1

\bibitem{Teitelboim:1983ux} 
C.~Teitelboim,
``Gravitation and Hamiltonian Structure in Two Space-Time Dimensions,''
Phys.\ Lett.\  {\bf 126B}, 41 (1983).
doi:10.1016/0370-2693(83)90012-6

\bibitem{Almheiri:2014cka} 
A.~Almheiri and J.~Polchinski,
``Models of AdS$_{2}$ backreaction and holography,''
JHEP {\bf 1511}, 014 (2015)
doi:10.1007/JHEP11(2015)014
[arXiv:1402.6334 [hep-th]].

\bibitem{Maldacena:2016upp} 
J.~Maldacena, D.~Stanford and Z.~Yang,
``Conformal symmetry and its breaking in two dimensional Nearly Anti-de-Sitter space,''
PTEP {\bf 2016}, no. 12, 12C104 (2016)
doi:10.1093/ptep/ptw124
[arXiv:1606.01857 [hep-th]].

\bibitem{Nayak:2018qej} 
P.~Nayak, A.~Shukla, R.~M.~Soni, S.~P.~Trivedi and V.~Vishal,
``On the Dynamics of Near-Extremal Black Holes,''
arXiv:1802.09547 [hep-th].


\end{thebibliography}
\end{document}